\documentclass[preprint,showpacs,preprintnumbers,amsmath,amssymb]{revtex4}

\usepackage{graphicx}
\usepackage{bm}

\begin{document}

\title{Inductive Detection of Magnetostrictive Resonance}
\author{Jang-ik Park, SangGap Lee and Insuk Yu\footnote{corresponding author: isyu@snu.ac.kr}}
\affiliation{School of Physics and Nano-Systems Institute
(NSI-NCRC), Seoul National University, Seoul 151-747, South Korea}

\author{Yongho Seo\footnote{corresponding author: yseo@sejong.ac.kr}}
\affiliation{Department of Nano Science \& Technology, Sejong
University, Seoul 143-747, Korea}

\date{\today}

\begin{abstract}
We have developed an inductive method to detect the
magnetostrictive resonance signal and applied it to an ultrasonic
magnetostrictive transducer sample. Slab shaped ferrite samples
are mounted in an RF coil and actuated by pulse modulated RF
magnetic field. A DC magnetic field is also applied and the
resonance signal from the sample is detected by the same coil
after the RF field is turned off. The detector system is similar
to a conventional pulse NMR system with quadrature detection. The
detected signal is sensitive to the bias DC field strength and
direction as well as the dimension of the sample.
\end{abstract}


\maketitle

\section{Introduction}

To study elastic properties of solid materials, a pulse resonance
technique composed of electromagnetic excitation and induction
signal detection has been used widely.\cite{Davis,Mason}.
Especially, for piezoelectric materials, Choi and Yu developed an
piezoelectric resonance technique to study ferroelectric
properties.\cite{Choi} For magnetic materials, a coupling between
electromagnetic field and sample magnetization mediates the
excitation and detection. Possible sources for this coupling are
the Lorentz force, the magnetization force, and the
magnetostriction force.\cite{Ogi1} The
magnetostriction\cite{Bozorth,Bozorth1,Slonczewski} phenomena
originated from elastic domain wall motion was used for an
oscillator first by Pierce\cite{Pierce}. This phenomenon has been
used for actuators and sensors by oscillating it and measuring the
frequency of elastic vibrations accurately. A representative
example is magnetostrictive transducers converting electrical
energy into mechanical energy. The magnetostriction transducers
have been applied for ultrasonic sound generators,
magnetostrictive optical wavelengths tuning, and acoustic delay
lines\cite{Hristoforou}. Recently, highly magnetostrictive
materials are investigated for practical application such as iron
rare-earth compounds of terfenol-d\cite{Chung} or ferromagnetic
oxide composites\cite{Chen}.

Lanotte {\it et al.}\cite{Lanotte0,Lanotte} had studied widely the
bias field dependency of the amplitude and frequency of the
acoustic waves in ferromagnetic materials using pulsed
electromagnetic excitation. They used exciting and pickup coils
with several hundred turns. The magnetoelastic wave amplitude was
measured as a function of the exciting pulse burst frequency in
the range of 2 $\sim$ 120 kHz. In a similar method, the
electromagnetic acoustic transducer (EMAT) technique has been used
for detecting acoustoelastic stress, attenuation coefficient,
grain size of metals, and magnetostriction
coefficients\cite{Kawashima,Hirao1,Hirao2,Ogi1,Ogi2}. For EMAT
technique, a ferromagnetic sample is located beneath of a
meander-line which is beneath of a permanent magnet or a disk
shape sample is located in a solenoid coil. The solenoid coil is
used for applying constant bias magnetic field along the sample
axis and the meander-line coil is to induce the dynamic field in
the circumferential direction through magnetostrictive effect and
receive the shear wave through the reversed magnetostrictive
effect.

In this paper, we report the rf pulse type resonance detection
method for magnetoacoustic wave in a slab shape ferrite material
used for an ultrasonic magnetostrictive transducer. The unique
features of our method are single impulse excitation and Fourier
transformation detection.

\section{Experiments}

Overall experimental setup is similar to a pulse type NMR
spectrometer and the sample is a commercially available
magnetostrictive transducer ferrite material commonly used for an
ultrasonic cleaner produced by TDK corp.\cite{TDK}

A schematics  of the experimental spectrometer setup used in our
experiment is shown in Fig. 1. This setup is similar to a
conventional pulsed NMR spectrometer with a series tuned sample
coil and quadrature detection receiver\cite{Abraham,Choi}. The
setup consists of the power amplifier which applies a pulse
modulated high voltage RF magnetic field to the sample inside the
sample coil and the quadrature detection receiver which picks up
the resonance signal from the sample. The sample and sample coil
are placed inside an electromagnet which apply the static magnetic
field to the sample. The signal from the sample is transmitted to
an analog-digital board in a PC with 8 bit voltage resolution and
1024pt digitization for each channel. The sample coil we used is a
solenoid of 25mm diameter and 10mm height in 11 turns tightly
wound. As shown in the Fig. 1, the dc bias magnetic field is
perpendicular to the pulsed RF magnetic field. When we applied the
static field parallel to the RF field, no signal was found.

The ferrite samples were cut into various sizes using a diamond
saw. The samples are slab shape of $\sim$ 1 mm thickness, $\sim$
10 mm width, and $\sim$ 50 mm length. A supporting structure made
by thin teflon sheet clamps one end of the sample. The sample is
rotatable by the teflon support. The spectrometer is tuned to 5
MHz and the pulses of width 0.5 $\mu$s and repetition time 10 ms
are applied throughout this work.

\section{Results and discussion}

The detected signal from the slab shape sample of 0.79 mm $\times$
9.06 mm $\times$ 42.85 mm is shown in Fig. 2. The signal is 64
times sampled to average out noise. The bias magnetic field
intensity is 0.6 kG and applied parallel to the sample surface.
Figure 2 shows the ring-down signal, when the sample is parallel
to magnetic field ($\theta = 0^{\circ}$). The inset of Fig. 2 is
the Fourier transformed (FT) waveform of the signal. The several
peaks in the spectrum are shown with equal spacing. The series of
peaks has frequencies
\begin{equation}
 f^{(n)} = {nc \over 2d},
 \label{fn}
\end{equation}
where $n$ is positive integer, $c$ acoustic wave velocity, $d$
length of the acoustic wave.\cite{Ogi2} The highest peak
corresponds to 12th harmonic ($f^{(12)} = 4.84$ MHz) and the
frequency spacing $\delta f = 0.403$ MHz.

We investigate the dependency of resonance frequency on the sample
geometry. The magnetic field is applied parallel to the sample
surface and kept to 0.6 kG. After the resonance signal of a sample
is measured, the sample is dismounted to narrow the width by
cutting it with a diamond saw. The frequency separation between
the peaks is shown as a function of sample width in Fig. 3(a). The
error bar comes from the fact that we have only 3 or 4 resonance
peaks in the spectrum. The graph shows the inverse proportionality
between the width $W$ of sample and the frequency separation
$\delta f$ of harmonics peaks, as

\begin{equation}
\delta f \propto {1 \over W}.
\end{equation}
Judging from this dependency, the acoustic waves are travelling in
width direction and possible mode is flexural mode. Therefore, $d$
in Eq. (\ref{fn}) is equal to $W$ and the acoustic wave velocity
$c$ is estimated as $7.31\times 10^3$ m/s.

Next we inspect the sample thickness dependence of the resonance
signal. Samples are prepared with 9.06 mm width, 42.85 mm length
and various thicknesses. The magnetic field is applied parallel to
the sample surface and kept to 0.6 kG, as the same as for previous
measurement. Fig. 3(b) shows the dependence of resonance frequency
on the sample thickness. Because the acoustic waves are travelling
in width direction, the frequency should not depend on thickness.
Real data shows a slight dependency on thickness. We attribute the
change of the resonance frequency to change of the internal strain
influenced by size effect.

The resonance frequency and amplitude are measured as functions
magnetic field $H$, as shown in Fig. 4 (a) and (b), respectively.
The dimension of this sample is 0.79 $\times$ 9.06 $\times$ 42.85
mm$^3$ and the bias field is applied parallel to the sample
surface. Among many peaks in Fourier transformed signal, the
highest peak frequency is shown in Fig. 4(a) The increase of the
resonance frequency depending on the bias field reflects the
increase of stiffness of the sample. The inset in Fig. 4(a) shows
the separation $\delta f$ between the harmonics of resonance
frequency versus the bias field dependence. The harmonics of
resonance frequency are measured in our measurable spectral range
between 4 MHz and 6 MHz. These data show a tendency that resonance
frequencies grows rapidly at low magnetic field and becomes
saturated in high magnetic field. This looks a typical behavior of
a magnetization change versus external field. Therefore, the
resonance frequency shift is affected by a mechanical strain
induced by the bias field. Similar behaviors on the other kinds of
magnetic samples were reported by Lanotte, {\it et al}, which was
explained as Young's modulus change ($\Delta E$
effect).\cite{Lanotte0,Lanotte} No noticeable hysteresis is found
in the resonance frequency change during increasing and decreasing
the bias magnetic field.

As shown in Fig. 4(b), the dependence of the amplitude on $H$
shows a peak behavior at $H=$ 0.7 kG. Similar behaviors were
reported in many literatures.\cite{Ogi1,Lanotte0,Thompson}
Thompson explained this peak behavior as magnetostrictive
contribution. We also attribute it to the field dependency of
magnetostriction coefficient. As $H$ increases, magnetic domains
begin to move against its stable configuration. When $H$ is larger
than its saturation field, the domain wall motion will disappear.
Therefore, for the saturation magnetic field, the $E$ and $f$ do
not change anymore and $A$ decreases down to zero.

The $f$ and $A$ are measured, as the bias field direction is
varied by rotating the teflon support holding the sample. The
sample dimension is the same as above and the bias magnetic field
is 0.6 kG. The resonance frequency and amplitude change versus the
angle ($\theta$ as shown in Fig. (1)) between the bias field and
sample surface are shown in Fig. 5(a) and (b), respectively. The
inset of the Fig. 5(a) shows the $\delta f$ between the harmonic
frequencies within our spectral range from 4 MHz to 6 MHz. This
experiment has no hysteresis as increasing or decreasing on angle
also. We attribute the angle dependency to the sample geometry.
Because the sample is a slab shape, it has magnetic anisotropy and
domain walls have preferential direction. The $\theta$ increase
can be considered as an effective magnetic field decrease.
Therefore, the $f$ and $A$ decrease as if they are changed by the
$H$ decrease as shown in Fig. 4.

In summary, we report a new experimental instrument using pulse
NMR spectrometer to study the magneto-elastic acoustic wave in
ferromagnetic materials. Our pulse modulated inductive detection
technique combined with Fourier transform analysis has an
advantage measuring the resonance spectrum at a single pulse
modulated rf excitation.

\acknowledgments

This work is supported by the National Core Research Center
program of the Korea Science and Engineering Foundation (KOSEF)
through the NANO Systems Institute of Seoul National University.


\newpage
\begin{figure}
\caption{Experimental setup is shown. The spectrometer is tuned at
5 MHz and has quadrature detection receiver. The ferrite sample is
located inside the electromagnet. The sample can be rotated
manually.}
\end{figure}

\begin{figure}
\caption{Ring down signal generated from a 0.79 mm $\times$ 9.06
mm $\times$ 42.85 mm sample. It is measured by an oscilloscope
with 1024 sampling points and 8-bit voltage resolution. The inset
shows the Fourier transform of the signal. In the Fourier
transform process, we exclude the first hundred points to reduce
the secondary effect from the preamplifier.}
\end{figure}

\begin{figure}
\caption{(a) The frequencies of harmonic resonance peaks depend on
the sample width. The spacing of frequencies is shown as a
function of the inverse of width. The samples are located parallel
to the bias field 0.6 kG. ($\theta = 0$.) (b) The dependency of
resonance frequency on the sample thickness is shown. This
dependency is due to the change of internal strain influenced by
size effect.}
\end{figure}

\begin{figure}
\caption{Bias field $H$ affects the resonance frequency $f$ (a)
and the signal amplitude $A$ (b). The inset shows the dependency
of $\delta f$ on the bias field.}
\end{figure}

\begin{figure}
\caption{The bias field direction on the sample affects the
resonance frequency (a) and the signal amplitude (b). $\theta$
indicates the angle between the sample and the bias field. The
inset shows the dependency of $\delta f$ on the field direction.}
\end{figure}
\end{document}